\documentclass{llncs}

\usepackage{stmaryrd}
\usepackage{enumerate}  
\usepackage{url}
\usepackage{graphicx}
\usepackage{amsmath}
\usepackage{amsfonts}
\usepackage{galois}
\usepackage{mathabx}
\usepackage{todonotes}
\usepackage{paralist}
\usepackage{multirow}
\usepackage{caption}
\usepackage{subfigure}
\usepackage{algorithmicx}
\usepackage[noend]{algpseudocode}
\usepackage{algorithm}
\usepackage{fancyvrb}
\usepackage{lscape}
\usepackage{listings}
\usepackage{pdflscape}
\usepackage{wrapfig}
\usepackage{tikz}
\usetikzlibrary{shapes}
\usetikzlibrary{arrows}
\usetikzlibrary{positioning}
\usetikzlibrary{automata}
\usetikzlibrary{calc}

\IfFileExists{./tmp/rm-margin.tex}{\input{tmp/rm-margin}}{}

\newtheorem{df}{Definition}
\newcommand{\program}{P}

\newcommand{\true}{\mathbf{tt}}
\newcommand{\false}{\mathbf{f}}

\newcommand{\Land}{\bigwedge}

\newcommand{\maps}{\rightarrow}

\newcommand{\union}{{\cup} }

\newcommand{\locationOf}[1]{loc(#1)}

\newcommand{\preds}{\pi}

\newcommand{\smem}{\mathcal{S}}
\newcommand{\tout}{TO}

\newcommand{\sa}{s_a}

\newcommand{\asp}{sp_a}

\newcommand{\concolic}{concolic}

\newcommand{\ourtool}{\textsc{Crabs}}

\newcommand{\crest}{\textsc{Crest}}
\newcommand{\synergy}{\textsc{Synergy}}
\newcommand{\cpachecker}{\textsc{CpaChecker}}
\newcommand{\fshell}{\textsc{Fshell}}
\newcommand{\svcomp}{\textsc{SvComp}}

\newtheorem{thm}{Theorem}

\newcommand{\rndbr}{RndBr}
\newcommand{\unfrnd}{UnfRnd}

\begin{document}

\title{Abstraction-driven Concolic Testing\thanks{This research was
    supported in part by the European Research Council (ERC) under
    grant 267989 (QUAREM) and by the Austrian Science Fund (FWF) under
    grants S11402-N23 (RiSE) and Z211-N23 (Wittgenstein Award).}  }

\author{Przemys\l aw Daca\inst{1} \and Ashutosh Gupta\inst{2} \and
  Thomas A.\ Henzinger\inst{1}} \institute{IST Austria, Austria \and
  Tata Institute for Fundamental Research, India}

\date{\today}

\maketitle
\begin{abstract}
Concolic testing is a promising method for generating test suites 
for large programs.
However, it suffers from the path-explosion problem and often fails to 
find tests that cover difficult-to-reach parts of programs.
In contrast, model checkers based on counterexample-guided abstraction
refinement explore programs exhaustively, while failing to scale on 
large programs with precision.
In this paper, we present a novel method that iteratively combines concolic 
testing and model checking to find a test suite for a given coverage criterion.
If concolic testing fails to cover some test goals, then the model checker 
refines its program abstraction to prove more paths infeasible, which reduces 
the search space for concolic testing.
We have implemented our method on top of the concolic-testing tool~\crest~and 
the model checker~\cpachecker.
We evaluated our tool on a collection of programs and a category of~\svcomp~benchmarks.
In our experiments, we observed an improvement in branch coverage compared to~\crest\
from $48\%$ to $63\%$ in the best case, and from $66\%$ to $71\%$ on average.


\end{abstract}

\section{Introduction}
\label{sec:intro}
%
Testing has been a corner stone of ensuring software reliability in
the industry, and despite the increasing scalability of software
verification tools, it still remains the preferred method for
debugging large software.
A test suite that achieves high code coverage is often required for
certification of safety-critical systems, for instance by the DO-178C
standard in avionics~\cite{rtca}.
Many methods for  automated test generation have been proposed
~\cite{Bird83,feedback,ARTOO,JPF,Symstra,FQL,Korat,verisoft}.
In the recent years, concolic testing has gained popularity as an
easy-to-apply method that scales to large programs.
Concolic testing~\cite{DART,SenMA05} explores program paths by a
combination of concrete and symbolic execution.
This method, however, suffers from the path-explosion problem and
fails to produce test cases that cover parts of programs that are
difficult to reach.

Concolic testing explores program paths using heuristic methods that select
the next path  depending on the paths explored so far.
Several heuristics for path exploration have been proposed that try to
maximize coverage of concolic
testing~\cite{BurnimS08,SAGE,Godefroid07}, e.g., randomly picking
program branches to explore, driving exploration toward uncovered
branches that are closest to the last explored branch, etc.
These heuristics, however, are limited by their ``local view'' of the
program semantics, i.e., they are only aware of the (in)feasibility of
the paths seen so far.
In contrast to testing, abstraction-based model checkers compute
abstract reachability graph of a program \cite{SLAM,BlastLazy02}.
The abstract reachability graph represents a ``global view'' of the
program, i.e., the graph contains all feasible paths.
Due to abstraction, not all paths contained in the abstract
reachability graph are guaranteed to be feasible, therefore abstract
model checking is not directly useful for generating test suites.

In this paper, we present a novel method to guide concolic testing by
an abstract reachability graph generated by a model checker.
The inputs to our method are a program and set of test goals, e.g.\
program branches or locations to be covered by testing.
Our method iterativly runs concolic testing and a 
counterexample-guided abstraction refinement (CEGAR) based
model checker~\cite{ClarkeCEGAR}.
The concolic tester aims to produce test cases covering as many goals
as possible within the given time budget.
In case the tester has not covered all the goals, the model checker is
called with the original program and the remaining uncovered goals
marked as error locations.
When the model checker reaches a goal, it either finds a test that
covers the goal or it refines the abstraction.
We have modified the CEGAR loop in the model checker such that it does
not terminate as soon as it finds a test, but instead it
removes the goal from the set of error locations and continues
building the abstraction.
As a consequence, the model checker refines the abstraction with respect
to the remaining goals.
After the model checker has exhausted its time budget, it returns
tests that cover some of the goals, and an abstraction.
The abstraction may prove that some of the remaining goals are unreachable,
thus they can be omitted by the testing process.
%
%

We further use the abstraction computed by the model checker to
construct a \emph{monitor}, which encodes the proofs of infeasibility
of some paths in the control-flow graph.
To this end, we construct a program that is an intersection of the
monitor and the program.
In the following iterations we run concolic testing on the intersected
program.
The monitor drives concolic testing away from the infeasible paths and
towards paths that  still may reach the remaining goals.
Due to this new ``global-view'' information concolic testing has fewer
paths to explore and is more likely to find test cases for the remaining
uncovered goals.
If we are still left with uncovered goals, the model checker is called
again to refine the abstraction, which further reduces the search
space for concolic testing.
Our method iterates until the user-defined time limit is reached.

The proposed method is configured by the ratio of time  spent on model
checking to the time spent on testing.
As we demonstrate in Section \ref{sec:example}, this ratio has a
strong impact on the test coverage achieved by our method.


We implemented our method in a tool called~\ourtool, which is built on
top of a \concolic-testing tool \crest~\cite{BurnimS08} and a
CEGAR-based model checker~\cpachecker~\cite{cpachecker}.
We applied our tool on three hand-crafted examples, three selected
published examples, and on $13$ examples from an~\svcomp~category. 
We compared our implementation with two tools: a concolic
tool~\crest~\cite{BurnimS08}, and a test-case generator~\fshell~based on
bounded model checking \cite{HolzerSTV08}.
The test objective was to cover program branches, and we calculate
test coverage as the ratio of branches covered by the generated test
suite to the number of branches that have not been proved unreachable.
For a time limit of one hour, our tool achieved coverage of $63\%$
compared to $48\%$ by other tools in the best case, and average coverage of $71\%$
compared to $66\%$ on the category examples.
In absolute numbers, our experiments may not appear very exciting.
However, experience suggests that in automated test generation
increasing test coverage by every $1\%$ becomes harder.
The experiments demonstrate that our method can cover branches that
are difficult to reach by other tools and, unlike most testing tools,
can prove that some testing goals are unreachable.

To summarize, the main contributions of the paper are:
\begin{itemize}
\item We present a novel configurable algorithm that iterativly
  combines concolic testing and model checking, such that concolic
  testing is guided by a program abstraction and the abstraction is
  refined for the remaining test goals.
\item We also present a modified CEGAR procedure that refines the abstraction
  with respect to the uncovered goals.
\item We provide an open-source tool~\cite{sourcepage} that implements
  the presented algorithm.
\item An experimental evaluation of our algorithm and comparison with
  other methods.
\end{itemize} 

The paper is organized as follows.
In Section~\ref{sec:example} we motivate our approach on examples. 
Section~\ref{sec:prelim} presents background notation and
concolic testing.
In Section~\ref{sec:cegar} we present our modified CEGAR procedure,
and in Section~\ref{sec:algo} we describe our main algorithm.
Finally,  Section~\ref{sec:exp} describes the experimental evaluation.


\section{Motivating Example}
\label{sec:example}
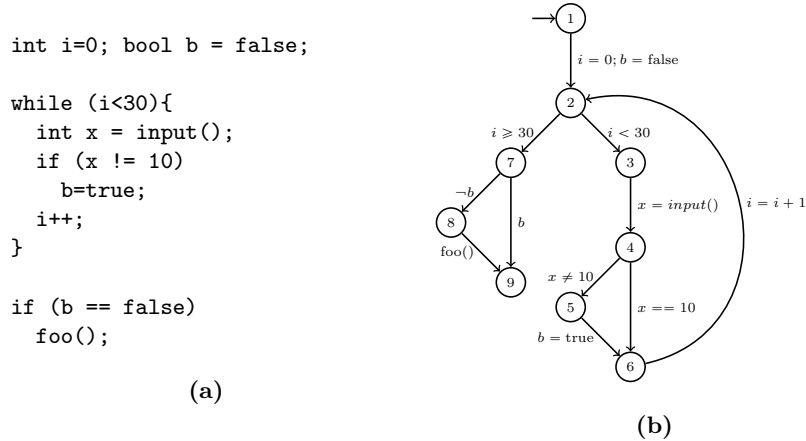
\begin{figure}[t]
  \centering
  \begin{minipage}{0.48\linewidth}
\begin{verbatim}
  int i=0; bool b = false;

  while (i<30){
    int x = input();
    if (x != 10)
      b=true;
    i++;
  }

  if (b == false) 
    foo();
\end{verbatim}
    \begin{center}
      {\bf (a)}
    \end{center}
  \end{minipage}
  \begin{minipage}{0.48\linewidth}
  \scalebox{0.75}{\begin{tikzpicture}[auto, node distance=1.5cm,->,initial text=, thick, align=left, font=\scriptsize]
\tikzstyle{state}=[circle, draw]
\tikzstyle{initstate}=[state,initial]
\tikzstyle{transition}=[->,>=stealth']

\node [initstate] (s1) {$1$};
\node [state,  below  of=s1] (s2) {$2$};
\node [state,  below right of=s2] (s3) {$3$};
\node [state,  below  of=s3] (s4) {$4$};

\node [state,  below left  of=s4] (s5) {$5$};
\node [state,  below right of=s5] (s6) {$6$};
\node [state,  below left of=s2] (s7) {$7$};
\node [state,  below left of=s7] (s8) {$8$};
\node [state,  below right of=s8] (s9) {$9$};

\path
        (s1) edge node[right] {$i=0; b=\mathrm{false}$} (s2)
        (s2) edge node[right] {$i<30$} (s3)
        (s2) edge node[left] {$i\geq30$} (s7)
        (s3) edge node[right] {$x=input()$} (s4)
        (s4) edge node[left] {$x\neq 10$} (s5)
        (s4) edge node[right] {$x==10$} (s6)
        (s5) edge node[left] {$b=\mathrm{true}$} (s6)
        (s6) edge[bend right, out=-90, in=-90, distance=3cm] node[right] {$i=i+1$} (s2)
        (s7) edge node[left] {$\neg b$} (s8)
        (s8) edge node[left] {$\mathrm{foo}()$} (s9)
        (s7) edge node[right] {$b$} (s9);

\end{tikzpicture}}    
  \begin{center}
    {\bf (b)}
  \end{center}
  \end{minipage}
  \caption{(a) A simple while program. (b) The control-flow graph of the program.}
\label{fig:while1}
\end{figure}


In this section, we illustrate effectiveness of our method on two
examples: a hand-crafted program, and a benchmark for worst-case
execution time analysis adapted from \cite{BanerjeeCR13}.

\paragraph{Simple loop}
In Figure \ref{fig:while1} we present a simple program with a single
while loop.
The program iterates $30$ times through the while loop, and in every
iteration it reads an input.
The test objective is to cover all locations of the program, in
particular to cover location $8$, where the library function
\texttt{foo()} is called.
To cover the call site to \texttt{foo()} the inputs in all iterations
must equal $10$, so only one out of $2^{30}$ ways to traverse the loop
covers \texttt{foo()}.
The standard concolic testing easily covers all locations, except for
\texttt{foo()} since it blindly explores exponentially many possible
ways to traverse the loop.
As a consequence, a concolic-testing tool is not able to generate a
complete test suite that executes \texttt{foo()} within one hour.

\begin{figure}[h]
  \centering
  \begin{minipage}[t]{0.48\linewidth}
    \centering
    \scalebox{0.75}{\begin{tikzpicture}[auto, node distance=1.5cm,->,initial text=, thick, align=left, font=\scriptsize]
\tikzstyle{state}=[rectangle, draw]
\tikzstyle{initstate}=[state,initial]
\tikzstyle{transition}=[->,>=stealth']

\node [initstate] (s1) {$1: true$};
\node [state,  below  of=s1] (s2) {$2: \neg b$};
\node [state,  below right of=s2] (s3) {$3: \neg b$};
\node [state,  below  of=s3] (s4) {$4: \neg b$};

\node [state,  below of=s4] (s6) {$6: \neg b$};
\node [state,  below left of=s2] (s7) {$7: \neg b$};
\node [state,  below  of=s7] (s8) {$8: \neg b$};
\node [state,  below of=s8] (s9) {$9: \neg b$};

\node [state,  below of=s6] (s2p) {$2: \neg b$};

\node [state,  right=1.3cm of s4] (t5) {$5: \neg b$};
\node [state,  below of=t5] (t6) {$6: b$};
\node [state,  below  of=t6] (t2) {$2: b$};
\node [state,  below right of=t2] (t3) {$3: b$};
\node [state,  below left of=t2] (t7) {$7: b$};
\node [state,  below of=t7] (t9) {$9: b$};
\node [state,  below  of=t3] (t4) {$4: b$};
\node [state,  below right of=t4] (t6p) {$6: b$};
\node [state,  below left of=t4,xshift=-7mm] (t5p) {$5: b$};
\node [state,  below of=t5p] (t6pp) {$6: b$};

\path
        (s1) edge node[right] {$i=0; b=\mathrm{false}$} (s2)
        (s2) edge node[right] {$i<30$} (s3)
        (s2) edge node[left] {$i\geq30$} (s7)
        (s3) edge node[right] {$x=input()$} (s4)
        (s4) edge node[below] {$x\neq 10$} (t5)
        (s4) edge node[left] {$x==10$} (s6)
        (s6) edge node[right] {$i=i+1$} (s2p)
        (s7) edge node[left] {$\neg b$} (s8)
        (s8) edge node[left] {$\mathrm{foo()}$} (s9)
        (s2p) edge[bend left, out=30, dashed] (s2)

        (t6) edge node[right] {$i=i+1$} (t2)
        (t2) edge node[right] {$i<30$} (t3)
        (t2) edge node[left] {$i\geq30$} (t7)
        (t3) edge node[left] {$x=input()$} (t4)
        (t4) edge node[left] {$x\neq 10$} (t5p)
        (t4) edge node[left] {$x==10$} (t6p)
        (t5) edge node[right] {$b=\mathrm{true}$} (t6)
        (t7) edge node[left] {$b$} (t9)
        (t5p) edge node[left] {$b=\mathrm{true}$} (t6pp)
        (t6p) edge[bend right, out=-15, in=-110, distance=2cm, dashed] (t6)
        (t6pp) edge[bend right, dashed] (t6p);
\end{tikzpicture}

}
    \\{\bf (a)}
  \end{minipage}
  \hspace{-6ex}
  \begin{minipage}[t]{0.49\linewidth}
  \centering
    \scalebox{0.75}{\begin{tikzpicture}[auto, node distance=1.5cm,->,initial text=, thick, align=left, font=\scriptsize]
\tikzstyle{state}=[circle, draw]
\tikzstyle{initstate}=[state,initial]
\tikzstyle{transition}=[->,>=stealth']

\node [initstate] (s1) {$1$};
\node [state,  below  of=s1] (s2) {$2$};
\node [state,  below right of=s2] (s3) {$3$};
\node [state,  below  of=s3] (s4) {$4$};

\node [state,  below of=s4] (s6) {$6$};
\node [state,  below left of=s2] (s7) {$7$};
\node [state,  below of=s7] (s8) {$8$};
\node [state,  below of=s8] (s9) {$9$};

\node [state,  right=1.3cm of s4] (t5) {$10$};
\node [state,  below of=t5] (t6) {$11$};
\node [state,  below  of=t6] (t2) {$12$};
\node [state,  below right of=t2] (t3) {$13$};
\node [state,  below left of=t2] (t7) {$14$};
\node [state,  below of=t7] (t9) {$15$};
\node [state,  below of=t3] (t4) {$16$};
\node [state,  below of=t4] (t17) {$17$};
\node [below=1cm of t4] (bogus) {};

\path
        (s1) edge node[right] {$i=0; b=\mathrm{false}$} (s2)
        (s2) edge node[right] {$i<30$} (s3)
        (s2) edge node[left] {$i\geq30$} (s7)
        (s3) edge node[right] {$x=input()$} (s4)
        (s4) edge node[below] {$x\neq 10$} (t5)
        (s4) edge node[right,yshift=-3mm] {$x==10$} (s6)
        (s6) edge[bend left, out=60] node[right,xshift=-1mm,yshift=3mm] {$i=i+1$} (s2)
        (s7) edge node[left] {$\mathrm{assume}(\neg b)$} (s8)
        (s8) edge node[left] {$\mathrm{foo()}$} (s9)

        (t6) edge node[right] {$i=i+1$} (t2)
        (t2) edge node[right] {$i<30$} (t3)
        (t2) edge node[left] {$i\geq30$} (t7)
        (t3) edge node[left] {$x=input()$} (t4)
       (t4) edge node[left] {$x\neq 10$} (t17)
        (t4) edge[bend right, out=-60, in=-120, distance=1.5cm]
        node[above,yshift=9mm,xshift=2mm] {$x==10$} (t6)
        (t17) edge[-,bend right, out=-60, in=-116, distance=2.5cm]
        node[above,yshift=9mm,xshift=2mm] {$b=true$} (t6)
        (t5) edge node {$b=\mathrm{true}$} (t6)
        (t7) edge node[left] {$\mathrm{assume}(b)$} (t9);

\end{tikzpicture}

}
    \\{\bf (b)}
  \end{minipage}
  \caption{(a) Abstraction refined with the predicate $b$. Dashed arrows
    show subsumption between abstract state. 
(b) The monitor obtained from the
    abstraction.
  }
  \label{fig:cfg-while-mon}
\end{figure}
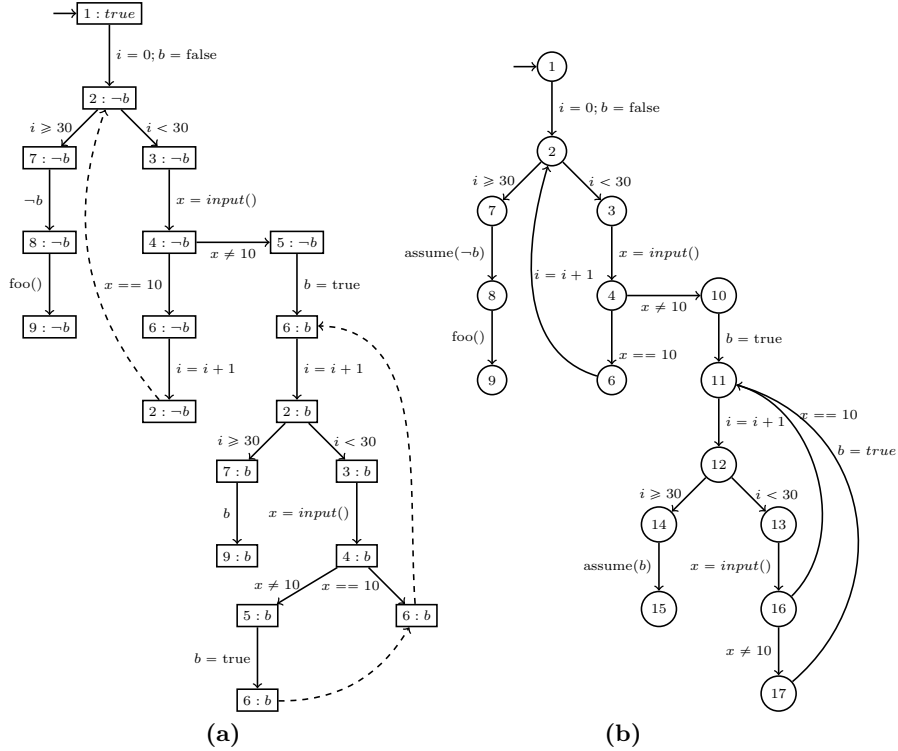

Our algorithm uses a concolic tester and a model checker based on
predicate abstraction, and runs them in alternation.
First, we run concolic tester on the example with a time budget of
$1$s.
As we have observed earlier, the concolic tester covers all locations of
the program except for \texttt{foo()}.
Then, we declare the call site to \texttt{foo()} as an error location
and call the model checker on the program for $5$s.
This time budget is sufficient for the model checker to perform only a
few refinements of the abstraction, without finding a feasible path
that covers \texttt{foo()}.  In particular, it finds an abstract
counterexample that goes through locations $1, 2, 3, 4, 5, 6, 2, 7, 8, 9$.
This counterexample is spurious, so the refinement procedure finds the
predicate ``$b$ holds.''
The abstraction refined with this predicate is showed in Figure
\ref{fig:cfg-while-mon}(a).

In the second iteration of the algorithm, we convert the refined
abstraction into a monitor $\mathcal{M}$ shown in Figure
\ref{fig:cfg-while-mon}(b).
A monitor is a control-flow graph that represents all the paths that
are allowed by the abstraction.
A monitor is constructed by removing subsumed states from the  abstraction.
We say that an abstract state $\sa$ is \emph{subsumed} by a state $\sa'$ if
$\sa=\sa'$, or $\sa'$ is more general than $\sa$.
To this end, the monitor includes all the abstract states that are not
subsumed and the edges between them.
The edges to the subsumed states are redirected to the states that subsume
them.

The monitor contains all the feasible paths of the program and is a
refinement of the control-flow graph of the original program.
Therefore, we may perform our subsequent concolic testing on the
monitor interpreted as a program.
In our example, the structure of the monitor in Figure
\ref{fig:cfg-while-mon}(b) encodes the information that \texttt{foo()}
can be reached only if $b$ is never set to true.
The refined control flow graph makes it easy for concolic testing to
cover the call to \texttt{foo()} --- it can simply backtrack whenever
the search goes to the part of the refined program where
\texttt{foo()} is unreachable.
Now, if we run $\crest$ on the monitor $\mathcal{M}$ then it finds the
test case in less than $1$s.

\paragraph{Nsichneu}
The ``nsichneu'' example is a benchmark for worst-case execution time
analysis \cite{Gustafsson:WCET2010:Benchmarks} and it simulates a
Petri net.
This program consists of a large number of if-then-else statements
closed in a deterministic loop.
The program maintains several integer variables and fixed-sized arrays
of integers.
These data objects are marked as \texttt{volatile} meaning that their
value can change at any time.
We made their initial values the input to the program.

The structure of this benchmark makes it challenging for many testing
techniques.
Testing based on bounded model checking (such as
\fshell\cite{HolzerSTV08}) unwinds the program up to a given bound and
encodes the reachability problem as a constraint-solving problem.
However, this method may not find goals that are deep in the program,
as the number of constraints grows quickly with the bound.
Test generation based on model checking \cite{Beyer04} also fails to
deliver high coverage on this example.
The model checker needs many predicates to find a feasible
counterexample, and the abstraction quickly becomes expensive to
maintain.
In contrast, pure concolic testing quickly covers easy-to-reach parts of the
program.
However, later it struggles to cover goals that are reachable by fewer
paths.

In our method, we run concolic testing and model checking
alternatively, each time with a time budget of $100$s.
Every iteration of model checking gives us a more refined monitor to
guide the testing process.
Initially, our approach covers goals at similar rate as pure concolic testing.
When the easy goals have been reached, our tool covers new goals
faster than concolic testing, due to the reachability information encoded in the
monitor, which allows the testing process to skip many long paths that
would fail to cover new goals. 
After one hour, our tool covers $63\%$ of the test goals compared to
$48\%$ by concolic testing.
\begin{wrapfigure}{r}{0.5\textwidth}
\vspace{-.5cm}
  \begin{center}
    \input{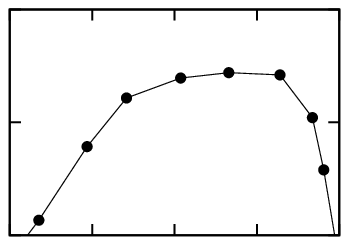}
    \caption{Test coverage vs. ratio of testing to total time in our method.}
    \label{fig:cov_plot_nsi}
  \end{center}
\vspace{-1cm}
\end{wrapfigure}


Furthermore, our method is configurable by the ratio of time spent on
model checking and concolic testing.
In Figure~\ref{fig:cov_plot_nsi} we present the effect of changing
this ratio on the example.
If we run only concolic testing then we obtain only $48\%$ coverage.
As we decrease the time spent on concolic testing, the coverage
increases up to $64\%$ and then starts decreasing.
On the other side of the spectrum, we generate tests by model checking
(as in \cite{Beyer04}) and obtain only $13.9\%$ coverage.
This observation allows one to configure our method for most effective testing
depending on the class of examples.


\section{Preliminaries}
\label{sec:prelim}
In this paper, we  consider only sequential programs and, for ease 
of presentation, we  consider programs without procedures.
Our method, however, is easily applicable on programs with procedures and 
our implementation supports them.

Let $V$ be a vector of variables names and
$V'$ be the vector of variables obtained by placing prime 
after each variable in $V$.
Let $F(V)$ be the set of first-order-logic formulas that only contain free variables from $V$.

\begin{df}[Program]
  A program $\program$ is a tuple $(V, Loc, \ell^I, E)$,
  where $V$ is a vector of variables,
  $Loc$ is a finite set of locations,
  $\ell^I \in Loc$ is the initial location,
  and $E \subseteq Loc \times F(V,V') \times Loc$ is a set of program transitions.
\end{df}
A \emph{control-flow graph} (CFG) is a graph representation of a program.
We define the \emph{product} of two programs $P_{i=1..2} = (V, Loc_i, \ell^I_i, E_i)$ as the program
$P_1\times P_2 = (V, Loc_1 \times Loc_2, (\ell^I_1, \ell^I_2), E)$, where
\[ E = \{ ((\ell_1, \ell_2), e, (\ell'_1, \ell'_2)) ~|~ (\ell_1, e, \ell'_1) \in E_1 \land (\ell_2, e, \ell'_2) \in E_2   \}.\]

A {\em guarded command} is a pair of a formula in $F(V)$ and 
a list of updates to variables in $V$.
For ease of notation, we may write the formula in a program transition as a guarded 
command over variables in $V$.
For example, let us consider $V = [x,y]$.
The formula represented by the guarded command $(x > y , [x := x + 1])$ is
$x > y \land x' = x + 1 \land y' = y$.
In our notation if a variable is not updated in the command
then the variable remains unchanged.
We use a special command $ variable := input()$ to model inputs to the
program, which logically means unconstrained update of the variable.
For example, the formula represented by the guarded command
$x := input()$ is $ y' = y$.
%
%
%
For an expression or formula $F$ we write $F[/i]$ to denote a formula
that is obtained after adding subscript $i+1$ to every primed variable
and $i$ to every unprimed variable.

A {\em valuation} is a mapping from the program variables $V$ to
values in the data domain.
%
A {\em state} $s = (l,v)$ consists of a program location $l$
and a valuation $v$.
For a state $s=(l,v)$ and a variable $x$, let $s(x)$ denote the
valuation of $x$ in $v$ and let $\locationOf{s}=l$.
A {\em path}  is a sequence $e_0,\dots,e_{n-1}$ of program transitions such that
$e_0 = (\ell^I,\_,\_)$,
and for $ 0 \leq i < n$, 
$e_i = (\ell_i, \_, \ell_{i+1}) \in E$. 
%
An {\em execution} corresponding to the path $e_0,\dots,e_{n-1}$ is a 
sequence of states $s_0 = (\ell_0,v_0),\dots
s_n = (\ell_n,v_n)$, such that 
1) $\ell_0 = \ell^I$, and
2)  for all $ 0 \leq i < n $, 
if $e_i = (\_,c_i(V,V'),\ell')$ then $\ell_{i+1} = \ell'$ and  $c_i(v_{i},v_{i+1})$ holds true.
%
%
We assume that for each execution of the program there exist exactly
one corresponding path, i.e., there is no non-determinism in the
program except inputs.

A path is represented symbolically by a set of path constraints, which
we define as follows.
Let $frame(x)$ be the formula $\Land_{y\in V: y \neq x} y' = y$.
Let $r_k$ be a variable that symbolically represent the $k$th input
on some path.
We assume the program does not contain any variable named $r$.
Let $e_0,\dots,e_{n-1}$ be a path.
If $e_i = (\_,[F,x := exp],\_)$ then let $C_i = (F \land x' = exp \land frame(x))[/i]$
and if $e_i = (\_,[F,x := input()],\_)$ then let $C_i = (F \land frame(x))[/i] \land x_{i+1} = r_k$,
where $r_0$ up to $r_{k-1}$ have been used in $C_0,\dots,C_{i-1}$.
The {\em path constraints} for the path is $C_0,\dots,C_{n-1}$.

A {\em test} of the program is a sequence of values.
A test $u_1,\dots,u_k$ {\em realizes} an execution 
$s_0,\dots,s_n$ and its corresponding path $e_0,\dots,e_{n-1} $ if  the
following conditions hold true:
\begin{itemize}
\item if $n = 0$, then $k=0$.
\item If $n > 0$ and $e_{n-1} = (\_, x:=input(),\_)$, $s_n(x) = u_k$ and $u_1,\dots,u_{k-1}$ realizes $s_0,\dots,s_{n-1}$.
\item Otherwise,  $u_1,\dots,u_k$ realizes $s_0,\dots,s_{n-1}$.
\end{itemize}
A path is said to be \emph{feasible} if there exists a test that realizes it.
In the above, we assume that the program does not read a variable
until its value is initialized within the program or explicitly taken
as input earlier.
Thus, the initial values are not part of tests.

In the context of test suit generation, we may refer to a transition as
a {\em branch} if the source location of the transition has multiple
outgoing transitions.
A test $t$ {\em covers}  branch $e$ if the test realizes 
a path that contains $e$.
Branch $e$ is {\em reachable} if there exists a test $t$ that covers $e$.
The \emph{test generation problem} is to find a set of tests that
covers every reachable branch in the program.

\subsection{Concolic~Testing}
\begin{algorithm}[t]
 \caption{\textsc{Concolic}$(\program = (V, L, \ell^I, E),G,t_b)$}
 \label{alg:concolic-testing}
\begin{algorithmic}[1]
   \Require {program $\program = (V, L, \ell^I,E)$, uncovered branches $G$, time budget $t_b$}
   \Ensure tests suite, uncovered branches

  \State $tst \gets ()$; 
  \State $\ell \gets \ell^I$; arbitrary $v$; $\smem \gets \lambda x\in V.\bot$ \Comment{initial values}
  \State $pathC \gets ()$; $suite \gets \emptyset$; $k=0$;
  \While{$ct < t_b$ and $G \neq \emptyset$} \Comment{$ct$ always has the current time}
        \If {$\exists e = (\ell,[F,x := exp],\ell')\in E$ such that $v \models F$} \Comment{expand}
                \State $G \gets G - \{e\}$; $\ell \gets \ell'$;
        	\State $pathC.push( F(\smem) )$ \Comment{$F(\smem)$ is substitution }
        	\If{ $exp = input()$}
                \If{ $|tst| = k$} $~w \gets randVal() $; $tst.push(w)$; {\bf else} $w \gets tst(k)$;
                \EndIf
                \State $v \gets v[x \mapsto w ]$;  $\smem \gets \smem[x\mapsto r_k]$; $k = k+1$
                \Else
                	\State $v \gets v[x \mapsto exp(v) ]$
                        \State $\smem \gets \smem[x \maps UpdateSymMem(\smem, exp, v)]$
                \EndIf
        \Else \Comment{backtrack}
        	\State $suite \gets suite \union \{tst\}$
        	\If {$\exists i < |pathC|$ such that  $\phi = \bigwedge_{j< i} pathC(j) \land \neg pathC(i)$ is sat}
                	\State $m = getModel(\phi)$
                        \State $l \gets $ number of distinct $r_i$s that occur in $\phi$
                	\State $tst \gets (m(r_0),\dots,m(r_{l-1}))$
                        \State {\bf goto} 2
                 \Else~{\bf break;}
                 \EndIf
        \EndIf
  \EndWhile
\State \Return $(suite,G)$\;
\end{algorithmic}
\end{algorithm}  



In concolic testing, a test suite is generated using both symbolic and
concrete execution. 
In Algorithm~\ref{alg:concolic-testing} we reproduce the procedure; the
presentation is modified such that we may use the procedure in our
main algorithm.
For simplicity of the presentation, we assume that there are 
at most two outgoing transitions at any program location and 
their guards are complementary to each other.
This assumption does not restrict the applicability of the
method.

The procedure takes a program $P = (V, Loc, \ell^I, E)$, a set of goal
branches $G$, and a time budget $t_b$ as input, and returns a test suite
that covers a subset of $G$ within the time budget $t_b$.
The procedure maintains a symbolic memory $\smem$, which is a partial
function from the program variables $V$ to symbolic expressions.
We use the symbol $\bot$ to denote an undefined value in a partial
function.
In addition, the procedure uses the following data structures: the
current location $\ell$, current valuation $v$ of variables, list
$pathC$ that contains constraints along the current path, test $tst$
that produces the current path, counter $k$ of inputs that have been
read on the current path, and a set $suite$ of tests seen so far.
We initialize all the collecting data structures to be empty,
 $\ell$ is initialized to be the initial location $\ell^I$, 
and the symbolic memory to be empty.

The algorithm proceeds by extending the current path by a transition
in each iteration of the while loop at line 4.
The loop runs until there are no goals to be covered or the procedure 
runs out of its time budget.
In the loop body, the condition checks if it is possible to
extend the current path by a transition $e = (\ell,[F, x := exp], \ell')$.
If the guard of $e$ satisfies the current valuation $v$ then $e$ is removed
from the set of goals and the current location is updated to $\ell'$.
In case $e$ has an input command $x:=input()$, then 1) the algorithm
updates $v(x)$ to the $k$th value from $tst$ if it is available, 2)
otherwise $v(x)$ is assigned a random value $w$, and $w$ is appended
to $tst$.
In either case, $\smem$ is updated by a fresh symbol $r_k$, assuming
$r_0$ to $r_{k-1}$ have been used so far.
If $e$ is not an input command, then both concrete and symbolic values
of $x$ are updated in $v$ and $\smem$ at line 10.

The symbolic memory is updated by the procedure $UpdateSymMem$.
$UpdateSymMem$ first computes $exp(\smem)$, and if the resulting
formula is beyond the capacity of available satisfiability checkers,
then it simplifies the formula by substituting the concrete values
from $v$ for some symbolic variables to make the formula decidable in
the chosen theory.
$UpdateSymMem$ is the key heuristics in concolic testing that
brings elements of 
concrete testing and symbolic execution together.
For details of this operation see~\cite{DART,SenMA05}.

At line 7, $pathC$ is extended by $F(\smem)$, which is the formula 
obtained after substituting every variable $x$ occurring 
in $F$ by $\smem(x)$.
We assume that variables are always initialized before usage, 
so $\smem$ is always defined for free variables in $F$.

In case the current path cannot be further extended, at lines 
16--19 the  procedure tries to find a branch on the path to backtrack.
For a chosen branch with index $i$, a formula is built that contains
the path constraints up to $i-1$ and the negation of the $i$th
constraint.
If this formula is satisfiable, then its model is converted to a new
test and path exploration restarts.
Note that the branch can be chosen non-deterministically, which allows
us to choose a wide range of heuristics for choosing the next path.
For example, the branch can be chosen at random or in the depth-first
manner by picking the largest unexplored branch $i$.
Another important heuristic that is implemented in~\crest~is 
to follow a branch that leads to the closest uncovered branch.


\section{Coverage-driven Abstraction Refinement}
\label{sec:cegar}
In this section, we present a modified version of CEGAR-based model checking 
that we use in our main algorithm.
Our modifications are: 1) the procedure continues until all goal
branches are covered by tests, proved unreachable or 
until the procedure reaches the time limit,
2) the procedure always returns an abstract reachability graph that
is closed under the abstract post operator.
%

The classical CEGAR-based model checking executes a program using an
abstract semantics, which is defined by an abstraction.
Typically, the abstraction is chosen such that the reachability graph
generated due to the abstract execution is finite.
If the computed reachability graph satisfies the correctness
specification, then the input program is correct.
Otherwise, the model checker finds an abstract counterexample, i.e.,
a path in the reachability graph that reaches an error state.
The abstract counterexample is spurious if there is no concrete
execution that corresponds to the abstract counterexample.
If the counterexample is not spurious then a bug has been found and
the model checker terminates.
In case of a spurious counterexample, the refinement procedure
refines the abstract model.
This is done by refining the abstraction to remove the
spurious counterexample, and the process restarts with the newly
refined abstraction.
After a number of iterations, the abstract model may have no more
counterexamples, which proves the correctness of the input program.
\begin{algorithm}[t]
\caption{\textsc{AbstractMC}($\program = (V, L, \ell^I, E)$,  $\pi$, $G$, $t_b$)}
 \label{alg:abstract-model-check}
 \begin{algorithmic}[1] 
   \Require {program $\program = (V, L, \ell^I,E)$, predicates $\preds$, uncovered branches $G$,\hspace{1.6cm}\mbox{}\hspace{2cm} time budget $t_b$}
   \Ensure tests, remaining branches, branches proved unreachable, new predicates,\hspace{1.6cm}\mbox{}
    abstract reachability graph
  \State $worklist~\gets~\{(\ell^I,\emptyset)\}$;
  $reach \gets \emptyset$;
  $subsume \gets \lambda \sa.\bot$; $parent((\ell_0,\emptyset))\gets \bot$
  \While{$worklist \neq \emptyset$}
  \State choose $(\ell,A) \in worklist$
  \State $worklist~\gets~worklist\setminus \{(\ell,A)\}$
  \If{$\textrm{false} \in A$ or $\exists \sa \in parent^*((\ell,A)).\; \sa \in sub$} 
     {\bf continue}
  \EndIf
  \State $reach~\gets~reach~\union~\{(\ell,A)\}$
  \If{$ \exists (\ell,A') \in reach-sub.\; A \subseteq A' $ }
  $subsume \gets subsume[ (\ell,A) \mapsto (\ell,A') ]$ 
  \Else
  \If{$ \exists (\ell,A') \in reach-sub.\; A' \subseteq A $ }
    $subsume \gets subsume [ (\ell,A') \mapsto (\ell,A) ]$
  \EndIf
  \For{  {\bf each } $e = (\ell,\rho,\ell') \in E $} 
  \State $A' \gets \asp(A,\rho)$;$worklist~\gets~worklist~\union~\{(\ell',A') \}$
  \State  $parent((\ell',A'))=(\ell,A)$; $trans((\ell',A'))=e$
  \If{$e \in G$} 
  \If{ $\exists m \models pathCons(\text{path to } (\ell',A'))$} 
  \State $G \gets G - \{e\}$
  \State $suite \gets suite \union \{ \text{the sequence of values of } r_k \text{s in }m\}$
  \Else
  \If{$ct < t_b$} \Comment{$ct$ has current time}
  \State $\pi \gets \pi \union$ \textsc{Refine}$( (\ell',A') )$;  {\bf goto} 1 
  \EndIf 
  \EndIf
  \EndIf
  \EndFor
  \EndIf
  \EndWhile 
  \State $U = G - \{e~|~\exists \sa \in reach.\; trans(\sa) = e\}$ \Comment{Unreachable goals}
  \State \Return{($suite$,$G-U$,$U$,$\pi$,$(reach,parent,subsume,trans)$)}
 \end{algorithmic}
\end{algorithm}  


In this paper, we use predicate abstraction for model checking.
Let $\preds$ be a set of predicates, which are formulas over variables
$V$.
We assume that $\preds$ always contains the predicate ``false.''
%
%
We define abstraction and concretization functions $\alpha$ and
$\gamma$ between the concrete domain of all formulas over $V$, and the
abstract domain of $2^{\preds}$:
$$
\alpha( \rho ) = \{ \varphi \in \preds ~|~ \rho \implies \varphi \}
\quad
\quad \gamma( A ) = \Land A,
$$
where $A\subseteq \pi$, and $\rho$ is a formula
over $V$.
An \emph{abstract state} $\sa$ of our program is an element of $Loc \times
2^{\preds}$.
Given an abstract state $(\ell, A)$ and a program transition $(\ell,
\phi, \ell')$, the abstract strongest post is defined as:
\[
\asp(A,\phi) =  \alpha( (\exists V.~\gamma(A)\land \phi(V,V'))[V'/V] ).
\]
The abstraction is refined by adding predicates to $\preds$.

In Algorithm~\ref{alg:abstract-model-check}, we present the
coverage-driven version of the CEGAR procedure.
We do not declare error locations or transitions, instead the procedure
takes goal transitions $G$ as input along with a program $P = (V, Loc,
\ell^I, E)$, predicates $\preds$, and a time budget $t_b$.
Reachable states are collected in $reach$, while $worklist$ contains
the frontier abstract states whose children are yet to be computed.
The procedure maintains functions $parent$ and $trans$, such that
if an  abstract state $\sa'$ is a child of a state $\sa$ by a transition $e$, 
then $parent(\sa')=\sa$ and $trans(\sa')=e$.
To guarantee termination, one needs to ensure that abstract states 
are not discovered repeatedly.
Therefore, the procedure also maintains the $subsume$ function, such that
$subsume((\ell,A)) = (\ell',A')$ only if 
$\ell=\ell'$ and $A \subseteq A'$.
We write $sub = \{ s ~|~ subsume(s) \neq \bot \}$ for the set of 
subsumed states.
We denote the reflexive transitive closure of $parent$ and $subsume$, by $parent^*$ and $subsume^*$, respectively. 

The algorithm proceeds as follows.
Initially, all collecting data structures are empty,
except $worklist$ containing the initial abstract state $(\ell^I,\emptyset)$.
The loop at line 2 expands the reachability graph
in every iteration.
At lines 3--4, it chooses an abstract state  $(l,A)$ from $worklist$.
If any ancestor of the state is already subsumed
or the state is false, the state is discarded and the next state is 
chosen.
Otherwise, $(l,A)$ is added to $reach$.
At lines 7--9, the $subsume$ function is updated.
Afterwords, if $(l,A)$ became subsumed then we proceed to choose
another state from $worklist$.
Otherwise, we create the children of $(l,A)$ in the loop at line 10 by 
the abstract post  $\asp$.
At line 12, $parent$ and $trans$ relations are updated.
%
At line 13, the procedure checks if the abstract reachability 
has reached any of the goal transitions.
If yes, then it checks the feasibility of the reaching path.
If the path is found to be feasible, we add the feasible 
solution as a test to the suite at line 16.
Otherwise, we refine and restart the reachability computation to
remove the spurious path from the abstract reachability at lines
18--19.
In case the algorithm has used its time budget, the refinement is not
performed, but the algorithm continues processing the states remaining
in $worklist$.
As a consequence, the algorithm always returns a complete abstract reachability graph.

We do not discuss details of the $\textsc{Refine}$ procedure.
The interested reader may read a more detailed exposition of CEGAR
in~\cite{Blast04}.

\paragraph{\bf Abstract reachability graph (ARG)}
The relations $parent$, $subsume$, and $trans$ together 
define an {\em abstract reachability graph (ARG)}, which is produced by 
\textsc{AbstractMC}.
A sequence of transitions $e_0,\ldots,e_{n-1}$ is a \emph{path in an ARG}
if there is a sequence of abstract state $s_0, \ldots, s_n \in reach$,
such that
\begin{enumerate}
\item $s_0 = (\ell^I, \emptyset)$,
\item for $1< i \leq n$ we have $parent(s_i)\in subsume^*(s_{i-1})$ and
$e_{i-1} = trans(s_i)$.
\end{enumerate}
\begin{thm}
  Every feasible path of the program $P$ is a path of an ARG.
  Moreover, every path in the ARG is a path of $P$.
\end{thm}

\textsc{AbstractMC} returns a set $suite$ of tests, set $G$ of
uncovered goals, proven unreachable goals $U$, set $\preds$ of
predicates, and the abstract reachability graph.
 
\paragraph{\bf Lazy abstraction}
Model checkers often implement various optimizations
in the computation of ARGs.
One of the key optimization is lazy abstraction \cite{BlastLazy02}.
CEGAR may learn many predicates that lead to ARGs that are expensive
to compute.
In lazy abstraction, one observes that not all applications of $\asp$ 
require the same predicates.
Let us suppose that the refinement procedure finds a new predicate
that {\em must} be added in specific place along a spurious
counterexample to remove this counterexample from future iterations.
In other paths, however, this predicate may be omitted.
This can be achieved by localizing predicates to parts of an ARG.
Support for lazy abstraction can easily be added by
additional data structures that record the importance of a 
predicate in different parts of programs.


\section{Abstraction-driven Concolic Testing}
\label{sec:algo}
In this section, we present our algorithm that combines
concolic testing and model checking.
The key idea is to use an ARG generated by a model checker to guide
concolic testing to explore more likely feasible parts of programs.

We start by presenting the function \textsc{MonitorFromARG} that converts
an ARG into a monitor program.
Let $\mathcal{A} = (reach,parent,subsume,trans)$ be an ARG.
The \emph{monitor of} $\mathcal{A}$ is defined as a program $\mathcal{M} =
(V, reach-sub, (\ell^I,\emptyset), E_1 \union E_2),$ where
\begin{itemize}
\item $E_1 = \{(\sa,e,\sa')~|~\sa = parent(\sa') \land e = trans(\sa') \land \sa' \not\in sub\}$,
\item $E_2 = \{(\sa,e,\sa'')~|~\exists \sa'.~\sa = parent(\sa') \land e = trans(\sa') \land \land \sa''\in~subsume^+(\sa') \land \sa''\not\in sub \}$.
\end{itemize}
The transitions in $E_1$ are due to the child-parent relation, when
the child abstract state is not subsumed.
In case the child state $\sa'$ is subsumed, then $E_2$ contains a
transition from the parent of $\sa'$ to the non-subsumed state $\sa''$ in 
$subsume^+(\sa')$, where $subsume^+$ denotes the transitive closure of $subsume$.
From the way we built an ARG, it follows that the state $\sa''$ is uniquely
defined and the monitor is always deterministic.

\begin{algorithm}[t]
  \caption{\ourtool($P = (V, Loc, \ell^i, E)$, $G$, $t_b$, $t_c$, $t_m$ )}
 \label{alg:our-alg}
\begin{algorithmic}[1]
\Require program $P = (V, Loc, \ell^i, E)$, branches $G \subseteq E$ to cover,
time budget for concolic testing $t_c$,
time budget for model checking $t_m$,
total time budget $t_b$,
\Ensure a test suite, set of provably unreachable branches
\State $\pi \gets \{\mathrm{false}\}$; $U \gets \emptyset$;\Comment{$U$ is a set of provably unreachable goals}
\State $suite \gets \emptyset$ \Comment{$suite$ is a set of test}
\State $\overline{P} \gets P$; $\overline{G} \gets G$ \Comment{program and goals for testing}
\State
\While {$G \neq \emptyset$ and $ ct < t_b$} \Comment{ $ct$ always has current time.}
\State $ (suite',\_) \gets \textsc{ConcolicTest}(\overline{P}, \overline{G}, ct+t_c)$
\State $ G \gets G - \{ g\in E ~|~ \exists tst\in suite'. tst\textrm{ covers } g\}$
\State $suite \gets suite \cup suite'$;
\If {$G \neq \emptyset$}
\State $(suite',G,U',\pi,\mathcal{A}) \gets \textsc{AbstractMC}(\program,\pi,G, ct+t_m)$ 
\State $suite \gets suite\cup suite'$; $U \gets U \union U'$
\State $\overline{P} \gets \overline{P} \times \textsc{MonitorFromARG}(\mathcal{A})$ \Comment{ see sec.~\ref{sec:algo} for MonitorFromARG}
\State $\overline{G} = \{ ((\ell,\_),e,(\ell',\_)) \in E_{\overline{P}} ~|~ (\ell, e,\ell')\in G  \}$
\EndIf
\EndWhile
\State \Return $(suite,U)$
\end{algorithmic}
\end{algorithm}


In Algorithm~\ref{alg:our-alg} we present our method~\ourtool.
\ourtool~takes as input a program $P$, a set $G$ of goal branches to be covered,
and time constraints: the total time limit $t_b$, and time budgets $t_c,t_m$ for a single iteration of concolic testing and model checking, respectively.
The algorithm returns a test suite for the covered goals, and a set of goals that
are provably unreachable.
The algorithm records in $G$ the set of remaining goals.
Similarly, $U$ collects the goal branches that
are proved unreachable by the model checker.
The algorithm maintains a set $\pi$ of predicates for abstraction, a program $\overline{P}$ for concolic testing, and a set $\overline{G}$ of goals for concolic testing.
The program $\overline{P}$ is initialized to the original program $P$, and in the following iterations becomes refined by the monitors.
The algorithm collects in $suite$ the tests generated by
concolic testing and model checking.

The program $\overline{P}$ is a refinement of the original program $P$, so a single goal branch in $P$ can map to many branches in the program $\overline{P}$.
For this reason, we perform testing for the set $\overline{G}$ of all possible extensions of $G$ to the branches in $\overline{P}$.
For simplicity, in our algorithm concolic testing tries to reach all goals in $\overline{G}$, even if they map to the same goal branch in $G$.
In the implementation, however, once concolic testing reaches a branch in $\overline{G}$, it removes all branches from $\overline{G}$ that have the same projection.

\ourtool~proceeds in iterations.
At line 6, it first runs concolic testing on the program
$\overline{P}$ and the goal branches $\overline{G}$ with the time
budget $t_c$.  The testing process returns tests $suite'$ and the
set of remaining branches.
Afterwords, if some branches remain to be tested, a model checker is
called on the program $P$ with predicates $\preds$, and a time budget
$t_m$ at line 10.
As we discussed in the previous section, the model checker builds an
abstract reachability graph (ARG), and produces tests if it finds
concrete paths to the goal branches.
Since the model checker runs for a limited amount of time,
it returns an abstract reachability graph that may have 
abstract paths to the goal branches, but no concrete paths
were discovered.
Moreover, if the ARG does not reach some goal branch then it is
certain that the branch is unreachable.
The model checker returns a new set $suite'$ of tests, remaining goals
$G$, and a set $U'$ of newly proved unreachable goals.
Furthermore, it also returns a new set $\preds$ of predicates for the
next call to the model checker, and an abstract reachability graph
$\mathcal{A}$.
At line 12, we construct a monitor from $\mathcal{A}$ by calling 
\textsc{MonitorFromARG}.
We construct the next program $\overline{P}$ by taking a product of the current $\overline{P}$ with the monitor.
We also update $\overline{G}$ to the set of all extensions of the branches in $G$ to the branches in $\overline{P}$.
In the next iteration concolic testing is called on $\overline{P}$,
which essentially explores the paths of $P$ that are allowed
by the monitors generated from the ARG.
The algorithm continues until it runs out of time budget
$t_b$ or no more goals remain.

The program $\overline{P}$ for testing is refined in every iteration by taking a
product with a new monitor.
This ensures that $\overline{P}$ always becomes more precise, even if the
consecutive abstractions do not strictly refine each other, i.e.
the ARG from iteration $i$ allows the set $\mathcal{L}$ of paths, while the
ARG from iteration $i+1$ allows the set $\mathcal{L'}$ such that
$\mathcal{L}'\not\subseteq \mathcal{L}$.
This phenomenon occurs when the model checker follows the lazy
abstraction paradigm, described in Section \ref{sec:cegar}.
In lazy abstraction, predicates are applied locally and some may be
lost due to refinement.
As a consequence, program parts that were pruned from an ARG may appear
again in some following ARG.
Another reason for this phenomenon may be a deliberate decision to
remove some predicates when the abstraction becomes too expensive to
maintain.




\section{Experiments}
\label{sec:exp}
We implemented our approach in a tool~\ourtool, built on top of the
concolic tester~\crest~\cite{BurnimS08} and the model
checker~\cpachecker~\cite{cpachecker}.
In our experiments, we observed an improvement in branch coverage
compared to~\crest\ from $48\%$ to $63\%$ in the best case, and from
$66\%$ to $71\%$ on average.
\paragraph{Benchmarks}
We evaluated our approach on a collection of programs: 1)~a set of
hand-crafted examples (listed in Appendix), 2)~example ``nsichneu"
\cite{Gustafsson:WCET2010:Benchmarks} described in
Section~\ref{sec:example} with varying number of loop iterations, 3)
benchmarks ``parport'' and ``cdaudio1'' from various categories of
~\svcomp~\cite{svcomp15}, 4)~all $13$ benchmarks from the
``ddv-machzwd''~\svcomp~category.

\paragraph{Optimizations}
Constructing an explicit product of a program and a monitor would be
cumbersome, due to complex semantics of the C language, e.g.\ the type
system and scoping rules.
To avoid this problem, our tool explores the product on-the-fly, by
keeping track of the program and monitor state.
We have done minor preprocessing of the examples, such that they can
be parsed by both~\crest~and~\cpachecker.
Furthermore, ~\cpachecker~ does not deal well with arrays, so in the
``nsichneu" example we replaced arrays of fixed size (at most $6$) by
a collection of variables.

\paragraph{Comparison of heuristics and tools}
We compare our tool with four other heuristics for guiding concolic
search that are implemented in~\crest: the depth-first search (DFS),
random branch search (\rndbr), uniform random search (\unfrnd), and
CFG-guided search; for details see \cite{BurnimS08}.  The depth-first
search is a classical way of traversing a tree of program paths.
In the random branch search, the branch to be flipped is chosen from
all the branches on the current execution with equal probability.
Similarly, in the uniform random search the branch to be flipped is
also picked at random, but the probability decreases with the position
of the branch on the execution.
In the CFG-guided heuristic the test process is guided by a distance
measure between program branches, which is computed statically on the
control-flow graph of the program.
This heuristic tries to drive exploration into branches that are
closer to the remaining test goals.
The concolic component of our tool uses the CFG-guided heuristic to
explore the product of a program and a monitor; this way branches
closer in the monitor are explored first.
In Appendix we show experimental results for our tool with the
other concolic heuristics, and we demonstrate that our approach
improves coverage for each of them.

We compared our approach with the tool \fshell~ \cite{HolzerSTV08},
which is based on the bounded model checker CBMC.  \fshell~unwinds the
control-flow graph until it fully explores all loop iterations and
checks satisfiability of paths that hit the testing goals.  This tool
does not return a test suite, unless all loops are fully explored.
%
%

\paragraph{Experimental setup}
All the tools were run with branch coverage as the test objective.
The coverage of a test suite is measured by the ratio $\frac{c}{r}$,
where $c$ is the number of branches covered by a test suite, and $r$
is the number of branches that have not been proved unreachable.
For ~\crest, we set $r$ to be the number of branches that are
reachable in the control-flow graph by graph search, which excludes
code that is trivially dead.
Our tool and~\crest~have the same number of test goals,
while~\fshell~counts more test goals on some examples.  We run our
tool in a configuration, where testing takes approximately $80\%$ of
the time budget.
All experiments were performed on a machine with an AMD Opteron $6134$
CPU and a memory limit of $12$GB, and were averaged over three runs.

\paragraph{Results}
The experimental evaluation  for a time budget of one hour is presented in Table \ref{tab:exp1h}; for more detailed results see Appendix.

After one hour, our tool achieved the highest coverage on most
examples.
The best case is ``nsichneu(17),'' where our tool achieved $63\%$
coverage compared to $48\%$ by the best other tool.
We show in Appendix, that if we run our tool with the DFS
heuristic, we obtain even higher coverage of $69\%$.
The hand-crafted examples demonstrate that our method, as well
as~\fshell, can reach program parts that are difficult to cover for
concolic testing.
In the benchmark category, our tool obtained average coverage of
$71\%$ compared to $66\%$ by~\crest.
In many examples, we obtain higher coverage by both reaching more
goals and proving that certain goals are unreachable.
~\fshell~generated test suites only for three examples, since on other
examples it was not able to fully unwind program loops.

\begin{landscape}
  \begin{table}[t]
    \centering
    \resizebox{\columnwidth}{!}{
  \begin{tabular}[a]{|lc||c|c|c|c|c|c|}
    \hline
    \multicolumn{2}{|c||}{Example} & {\ourtool-CFG~(this paper)} & \crest-DFS\cite{DART,SenMA05}& \crest-CFG\cite{BurnimS08} & \crest-\unfrnd\cite{BurnimS08} & \crest-\rndbr\cite{BurnimS08} & \fshell\cite{HolzerSTV08} \\
name		& branches	& coverage	& coverage	& coverage	& coverage	& coverage	& coverage \\
\hline
simple-while 	&  12   & 12/12 (100\%)		& 11/12 (91.2\%)	& 11/12 (91.2\%)	& 11/12 (91.2\%)	& 11/12 (91.2\%) 	& 12/12 (100\%)	\\
branches 	&  12   & 12/12 (100\%)	  	& 9/12 (75\%)		& 7/12 (58.3\%)		& 12/12 (91.2\%)	& 12/12 (91.2\%)	& 12/12 (100\%) \\
unreach  	& 10 	&    9/9 (100\%)	& 9/10 (90\%) 		& 9/10 (90\%) 		& 9/10 (90\%) 		& 9/10 (90\%) 		& 9/9 (100\%) \\
\hline
nsichneu(2)    	& 5786 & 3843/5753 (66.8\%) 	& \textbf{5365/5786 (92.7\%)} & 3098/5786 (53.5\%) & 2559/5786 (44.2\%) & 2196/5786 (38.0\%) 	&    4520/5786 (78.1\%) \\
nsichneu(9)    	& 5786  &3720/5756 (64.6\%)   	& \textbf{4224/5786 (73.0\%)} & 2843/5786 (49.1\%) & 2493/5786 (43.1\%) & 2187/5786 (37.8\%) 	& 1261/5786 (21.8\%)  	\\
nsichneu(17)   	& 5786 & \textbf{3619/5746 (63.0\%)} & 2086/5786 (36.1\%) & 2758/5786 (47.7\%) 	& 2476/5786 (42.8\%) & 2161/5786 (37.3\%) 	& \tout    		\\
parport		& 920 & \textbf{215/598 (35.9\%)}	& 215/920 (23.4\%) 	& 215/920 (23.4\%)  	& 215/920 (23.4\%)	& 215/920 (23.4\%)	& \tout			\\
cdaudio1 	& 340 & 248/249 (99.6\%) 	& 250/340 (73.5\%) 	& 250/340 (73.5\%) 	& 246/340 (72.3\%) 	& 250/340 (73.5\%) 	& \textbf{266/266 (100\%)}  \\
\hline
ddv\_outb      &  206 & \textbf{ 137/194  (70.8\%)} &   78/206  (37.9\%) &  136/206  (66.2\%) &  111/206  (54.2\%) &  135/206  (65.7\%) &    \tout       \\
ddv\_pthread   &  200 & \textbf{ 134/189  (71.3\%)} &   73/200  (36.7\%) &  131/200  (65.5\%) &  109/200  (54.8\%) &  130/200  (65.2\%) &    \tout       \\
ddv\_outwp     &  200 & \textbf{ 134/189  (70.8\%)} &   73/200  (36.7\%) &  131/200  (65.5\%) &  107/200  (53.8\%) &  129/200  (64.7\%) &    \tout       \\
ddv\_allfalse  &  214 & \textbf{ 143/199  (72.0\%)} &   83/214  (38.9\%) &  141/214  (66.0\%) &  123/214  (57.6\%) &  140/214  (65.6\%) &    \tout       \\
ddv\_inwp      &  200 & \textbf{ 134/189  (70.7\%)} &   76/200  (38.2\%) &  131/200  (65.5\%) &  108/200  (54.2\%) &  129/200  (64.8\%) &    \tout       \\
ddv\_inbp      &  200 & \textbf{ 133/189  (70.3\%)} &   73/200  (36.5\%) &  130/200  (65.3\%) &  109/200  (54.5\%) &  130/200  (65.2\%) &    \tout       \\
ddv\_outlp     &  200 & \textbf{ 133/189  (70.1\%)} &   73/200  (36.5\%) &  130/200  (65.2\%) &  109/200  (54.7\%) &  130/200  (65.2\%) &    \tout       \\
ddv\_outbp     &  200 & \textbf{ 134/188  (71.2\%)} &   73/200  (36.5\%) &  130/200  (65.0\%) &  106/200  (53.3\%) &  130/200  (65.0\%) &    \tout       \\
ddv\_inl       &  200 & \textbf{ 134/190  (70.7\%)} &   89/200  (44.8\%) &  131/200  (65.8\%) &  109/200  (54.7\%) &  129/200  (64.8\%) &    \tout       \\
ddv\_inlp      &  200 & \textbf{ 134/189  (71.1\%)} &   75/200  (37.8\%) &  131/200  (65.5\%) &  108/200  (54.3\%) &  129/200  (64.8\%) &    \tout       \\
ddv\_inw       &  206 & \textbf{ 139/194  (72.0\%)} &   80/206  (39.2\%) &  136/206  (66.3\%) &  112/206  (54.5\%) &  135/206  (65.7\%) &    \tout       \\
ddv\_inb       &  200 & \textbf{ 133/189  (70.5\%)} &   73/200  (36.5\%) &  130/200  (65.3\%) &  114/200  (57.3\%) &  130/200  (65.2\%) &    \tout       \\
ddv\_outl      &  200 & \textbf{ 133/189  (70.5\%)} &   73/200  (36.5\%) &  131/200  (65.5\%) &  109/200  (54.8\%) &  131/200  (65.7\%) &    \tout       \\
\hline
  \end{tabular}}
  \caption{Experimental results for one hour.~\rndbr~stands for ``random branch search'' and~\unfrnd~for ``uniform random search.''~\tout~ means that no suite was generated before the time limit.}
  \label{tab:exp1h}
\end{table}
\end{landscape}



\section{Related Work}
\label{sec:related}
Testing literature is  rich, so we only highlight the most prominent approaches. 
Random testing \cite{feedback,ARTOO,Bird83} can cheaply cover shallow
parts of the program, but it may quickly reach a plateau where
coverage does not increase.
Another testing method is to construct symbolic objects that represent
complex input to a program \cite{JPF,Symstra}.
In \cite{Korat} objects for testing program are systematically
constructed up to a given bound.  The approach of \cite{verisoft}
tests a concurrent program by exploring schedules using partial-order
reduction techniques.

Concolic testing suffers from the path-explosion problem, so various
search orders testing have been proposed, several of them are
discussed in Section \ref{sec:exp}.  
In \cite{SAGE} multiple input vectors are generated from a single
symbolic path by negating constraints on the path one-by-one, which
allows the algorithm to exercise paths at different depths of the
program.
Hybrid concolic testing \cite{Majumdar07} uses random testing to
quickly reach deep program statements and then concolic testing to
explore the close neighborhood of that point.
%

%
Our work is closest related to \textsc{Synergy}
\cite{Synergy,BeckmanNRS08,GodefroidNRT10}.
\textsc{Synergy} is an approach for verification of safety properties that
maintains a program abstraction and a forest of tested paths.
Abstract error traces are ordered such that they follow some
tested execution until the last intersection with the forest.
If an ordered abstract trace is feasible, then a longer concrete
path is added to the forest; otherwise, the abstraction is
refined.
Compared to \textsc{Synergy} our method has several key differences.
%
%
First, in~\synergy~model checking and test generation work as a single
process, while in our approach these components are independent and
communicate only by a monitor.
%
%
Second, unlike us, \synergy~does not pass the complete abstract model of the
program to concolic testing, where the testing heuristics guides the
search.
Finally, in our approach we can configure the ratio of model checking
to testing, while in~\synergy~every unsuccessful execution leads to
refinement.
%
%
%

Another related work is \cite{TandemETH-15}, where concolic testing is
guided towards program parts that a static analyzer was not able to
verify. In contrast to our approach, the abstraction is not refined.
In \cite{JCW2015} conditional model checking is used to generate a
residual that represents the program part that has been left
unverified; the residual is then tested.

The work of \cite{RungtaMV09} applies program analysis to identify
control locations in a concurrent program that are relevant for
reaching the target state.
These locations guide symbolic search toward the target and predicates
in failed symbolic executions are analyzed to find new relevant
locations.
The \textsc{Check'n'Crash} \cite{CsallnerS05} tool uses a constraint
solver to reproduce and check errors found by static analysis of a
program.
In \cite{CsallnerSX08} the precision of static analysis was improved
by adding a dynamic invariant detection .

The algorithm of \cite{Latest} presents a testing method, where a
program is simplified by replacing function calls by unconstrained
input.
%
Spurious counterexample are removed in a CEGAR loop by lazily
inserting function bodies.
In contrast, our method performs testing on a concrete program and
counterexamples are always sound.

A number of papers consider testing program abstraction with bounded
model checking (BMC).
If the abstraction is sufficiently small, then a program invariant can
be established by exhaustively testing the abstraction with BMC.
In \cite{Kroening06} a Boolean circuit is abstracted, such that it
decreases the bound that needs to be explored in an exhaustive BMC
search.
In \cite{GuptaS05} BMC is run on an abstract model up to some bound.
If the invariant is not violated, then the model is replaced by an unsat
core and the bound is incremented.
If a spurious counterexample is found, then clauses that appear in the unsat core are added to the abstraction.
%




\section{Conclusion}
\label{sec:conc}
We presented an algorithm that combines model checking and 
concolic testing synergistically.
Our method iterativly runs concolic
testing and model checking, such that concolic testing is guided by a program
abstraction, and the abstraction is refined for the remaining test goals.
Our experiments demonstrated that the presented method can increase branch
coverage compared to both concolic testing, and test
generation based on model checking.

We also observed that our method is highly sensitive to
optimizations and heuristics available in the model checker.
For instance, lazy abstraction allows the model checker to get pass 
bottlenecks created due to over-precision in some 
parts of ARGs.
However, lazy abstraction may lead to a monitor
that is less precise than the monitors of the past iterations, which
 may lead to stalled progress in covering new goals
by our algorithm.
In the future work, we will study such complimentary 
effects of various heuristics in model checkers to
find the optimal design of model checkers to assist
a concolic-testing tool.
We believe that adding this feature will further improve
the coverage of our tool.


\subsubsection*{Acknowledgments}
We thank Andrey Kupriyanov for feedback on the manuscript, and
Michael Tautschnig for help with preparing the experiments.

\clearpage
\bibliographystyle{abbrv}
\bibliography{biblio}

\clearpage
\appendix
\section*{Appendix}
\subsection*{Experiments for concolic search orders}
\label{sec:exp_d}

Our approach is independent of the search order used in concolic testing.
We modified the three search orders in~\crest~to work with a monitor: DFS, CFG and~\rndbr.
The experimental evaluation is shown in  Tables \ref{tab:exp_d} and \ref{tab:exp1h_d}.
The results suggest that using  information from a monitor increases test coverage for all three heuristics.

\begin{landscape}
  \begin{table}[h]
    \centering
    \resizebox{\columnwidth}{!}{
      \begin{tabular}[a]{|lc||c|c||c|c||c|c|}
        \hline
        \multicolumn{2}{|c||}{Example} & {\ourtool-DFS~(this paper)} & \crest-DFS\cite{DART,SenMA05}& {\ourtool-CFG~(this paper)} &\crest-CFG\cite{BurnimS08}  & {\ourtool-\rndbr~(this paper)} & \crest-\rndbr\cite{BurnimS08} \\
        name		& branches	& coverage	& coverage	& coverage	& coverage	& coverage	& coverage 	\\
        \hline
nsichneu(2)     & 5786 & \textbf{5329/5765 (92.4\%)} & 5264/5786 (91.0\%) & 3351/5763 (58.1\%) & 2807/5786 (48.5\%) &  2344/5756 (40.7\%) & 2143/5786 (37.0\%)  \\
nsichneu(9)     & 5786 & \textbf{3332/5771 (57.7\%)} & 2086/5786 (36.1\%) & 2978/5761 (51.7\%) & 2656/5786 (45.9\%) &  2269/5764 (39.4\%) & 2128/5786 (36.8\%)  \\
nsichneu(17)      & 5786 & \textbf{2949/5776 (51.1\%)} & 1453/5786 (25.1\%) & 2842/5764 (49.3\%) & 2546/5786 (44.0\%) &  2305/5758 (40.0\%) & 2117/5786 (36.6\%)  \\
\hline
ddv\_allfalse  &  214 &   87/208  (42.0\%) &   83/214  (38.9\%) & \textbf{ 142/205  (69.3\%)} &  134/214  (62.9\%) &  139/203  (68.8\%) &  131/214  (61.5\%)  \\
ddv\_outlp     &  200 &   81/195  (41.4\%) &   73/200  (36.5\%) & \textbf{ 133/189  (70.1\%)} &  123/200  (61.7\%) &  125/191  (65.7\%) &  121/200  (60.7\%)  \\
ddv\_inl       &  200 &   76/194  (39.4\%) &   89/200  (44.8\%) & \textbf{ 132/189  (69.9\%)} &  126/200  (63.3\%) &    130/192  (67.5\%) &  119/200  (59.8\%)  \\
ddv\_pthread   &  200 &   76/195  (39.1\%) &   73/200  (36.7\%) & \textbf{ 132/189  (69.9\%)} &  122/200  (61.0\%) &    130/190  (68.7\%) &  122/200  (61.0\%)  \\
ddv\_outl      &  200 &   84/194  (43.6\%) &   73/200  (36.5\%) & \textbf{ 132/190  (69.7\%)} &  117/200  (58.7\%) &    130/189  (68.8\%) &  125/200  (62.5\%)  \\
ddv\_inlp      &  200 &   77/194  (39.9\%) &   73/200  (36.7\%) & \textbf{ 132/190  (69.7\%)} &  122/200  (61.0\%) &    129/190  (67.9\%) &  121/200  (60.5\%)  \\
ddv\_outwp     &  200 &   84/194  (43.6\%) &   73/200  (36.7\%) & \textbf{ 133/190  (69.9\%)} &  122/200  (61.2\%) &    129/188  (68.7\%) &  124/200  (62.0\%)  \\
ddv\_inbp      &  200 &   88/194  (45.4\%) &   73/200  (36.5\%) & \textbf{ 132/190  (69.7\%)} &  120/200  (60.0\%) &    129/189  (68.4\%) &  120/200  (60.2\%)  \\
ddv\_outbp     &  200 &   76/194  (39.2\%) &   73/200  (36.5\%) &  117/189  (62.0\%) &  119/200  (59.8\%) &   \textbf{ 131/190  (69.2\%)} &  119/200  (59.5\%)  \\
ddv\_inw       &  206 &   83/200  (41.5\%) &   80/206  (39.2\%) & \textbf{ 137/195  (70.3\%)} &  124/206  (60.5\%) &     136/196  (69.6\%) &  125/206  (60.7\%)  \\
ddv\_outb      &  206 &   87/199  (43.7\%) &   78/206  (37.9\%) & \textbf{ 136/195  (69.8\%)} &  130/206  (63.4\%) &    131/196  (67.1\%) &  129/206  (62.8\%)  \\
ddv\_inwp      &  200 &   83/195  (42.5\%) &   76/200  (38.2\%) & \textbf{ 132/189  (70.1\%)} &  122/200  (61.2\%) &    129/191  (67.5\%) &  118/200  (59.2\%)  \\
ddv\_inb       &  200 &   85/194  (44.2\%) &   73/200  (36.5\%) & \textbf{ 131/189  (69.7\%)} &  121/200  (60.5\%) &    128/189  (67.7\%) &  121/200  (60.8\%)  \\
\hline
      \end{tabular}}
    \caption{Detailed experimental results for $15$m.}
    \label{tab:exp_d}
    \resizebox{\columnwidth}{!}{
      \begin{tabular}[a]{|lc||c|c||c|c||c|c|}
        \hline
        \multicolumn{2}{|c||}{Example} & {\ourtool-DFS~(this paper)} & \crest-DFS\cite{DART,SenMA05}& {\ourtool-CFG~(this paper)} &\crest-CFG\cite{BurnimS08} &  {\ourtool-\rndbr~(this paper)} & \crest-\rndbr\cite{BurnimS08} \\
        name		& branches	& coverage	& coverage	& coverage	& coverage	& coverage	& coverage 	\\
\hline
nsichneu(2)     & 5786 & \textbf{5406/5758 (93.9\%)} & 5365/5786 (92.7\%) & 3843/5753 (66.8\%) & 3098/5786 (53.5\%) &  2565/5738 (44.7\%) & 2196/5786 (38.0\%)  \\
nsichneu(9)     & 5786 & 4042/5755 (70.2\%) & \textbf{4224/5786 (73.0\%)} & 3720/5756 (64.6\%) & 2843/5786 (49.1\%) &  2474/5742 (43.1\%) & 2187/5786 (37.8\%)  \\
nsichneu(17)      & 5786 & \textbf{3951/5761 (68.6\%)} & 2086/5786 (36.1\%) & 3619/5746 (63.0\%) & 2758/5786 (47.7\%) &  2460/5746 (42.8\%) & 2161/5786 (37.3\%)  \\
\hline
ddv\_allfalse  &  214 &   87/207  (42.2\%) &   83/214  (38.9\%) & \textbf{ 143/199  (72.0\%)} &  141/214  (66.0\%) &    142/199  (71.4\%) &  140/214  (65.6\%)  \\
ddv\_outlp     &  200 &   78/193  (40.7\%) &   73/200  (36.5\%) & \textbf{ 133/189  (70.1\%)} &  130/200  (65.2\%) &    132/188  (70.0\%) &  130/200  (65.2\%)  \\
ddv\_inl       &  200 &   80/193  (41.7\%) &   89/200  (44.8\%) & \textbf{ 134/190  (70.7\%)} &  131/200  (65.8\%) &    132/188  (70.0\%) &  129/200  (64.8\%)  \\
ddv\_pthread   &  200 &   78/191  (41.1\%) &   73/200  (36.7\%) & \textbf{ 134/189  (71.3\%)} &  131/200  (65.5\%) &    132/190  (69.5\%) &  130/200  (65.2\%)  \\
ddv\_outl      &  200 &  102/192  (53.1\%) &   73/200  (36.5\%) & \textbf{ 133/189  (70.5\%)} &  131/200  (65.5\%) &    131/189  (69.6\%) &  131/200  (65.7\%)  \\
ddv\_inlp      &  200 &   79/193  (40.9\%) &   75/200  (37.8\%) & \textbf{ 134/189  (71.1\%)} &  131/200  (65.5\%) &    133/188  (70.6\%) &  129/200  (64.8\%)  \\
ddv\_outwp     &  200 &   82/191  (42.9\%) &   73/200  (36.7\%) & \textbf{ 134/189  (70.8\%)} &  131/200  (65.5\%) &    131/188  (69.9\%) &  129/200  (64.7\%)  \\
ddv\_inbp      &  200 &   90/191  (47.5\%) &   73/200  (36.5\%) & \textbf{ 133/189  (70.3\%)} &  130/200  (65.3\%) &    132/189  (69.8\%) &  130/200  (65.2\%)  \\
ddv\_outbp     &  200 &   77/192  (40.1\%) &   73/200  (36.5\%) & \textbf{ 134/188  (71.2\%)} &  130/200  (65.0\%) &    132/188  (70.3\%) &  130/200  (65.0\%)  \\
ddv\_inw       &  206 &   91/198  (46.0\%) &   80/206  (39.2\%) & \textbf{ 139/194  (72.0\%)} &  136/206  (66.3\%) &    137/194  (70.4\%) &  135/206  (65.7\%)  \\
ddv\_outb      &  206 &   94/195  (48.2\%) &   78/206  (37.9\%) &  137/194  (70.8\%) &  136/206  (66.2\%) &   \textbf{ 137/193  (71.2\%)} &  135/206  (65.7\%)  \\
ddv\_inwp      &  200 &   87/191  (45.5\%) &   76/200  (38.2\%) & \textbf{ 134/189  (70.7\%)} &  131/200  (65.5\%) &    132/188  (70.0\%) &  129/200  (64.8\%)  \\
ddv\_inb       &  200 &   82/192  (42.8\%) &   73/200  (36.5\%) & \textbf{ 133/189  (70.5\%)} &  130/200  (65.3\%) &    132/189  (69.8\%) &  130/200  (65.2\%)  \\

\hline
  \end{tabular}}
  \caption{Detailed experimental results for  $1$h.}
  \label{tab:exp1h_d}
\end{table}
\end{landscape}


\subsection*{Hand-craft program}
\label{sec:hc}
We list the hand-craft examples that were used for experimental evaluation.

\begin{figure}[h]
  \centering
\begin{verbatim}
const int B=30;

int main(void) {
  int i=0;
  int x;
  int b = 0;
  while (i<B){
    x = input();
    // we distinguish two cases
    // so that corner cases of DFS
    //don't give full coverage
    if (i==6 && x== 4){b=1;}

    if (i!=6 && x!= 4){b=1;}

    i++;
  }
  if (b==0) {printf("Goal!\n");}  
}
\end{verbatim} 
  \caption{The ``simple-while'' example.}
\label{fig:while}
\end{figure}

\begin{figure}[h]
  \centering
\begin{verbatim}
const int N=30;

// function that takes long time to analyze
int foo(int y){
  int i,c=0;
  for (int j=0;j<N;j++){
    i = input();
    if (i == c)
      c++;
  }
  return c;
}

void bar(int y){
  if (y<20){
    printf("Goal one\n");
  } else {
    printf("Goal two\n");
  }
}

int main(void) {
  int y,x;
  y = input();
  if (y>0){
    x = foo(y);
  } else {
    x = foo(y+1);
  }
  if (x==5 && y>10){
    // discover that can be reached only if y>0
    bar(y);
  }
}
\end{verbatim} 
  \caption{The ``branches'' example.}
\label{fig:branches}
\end{figure}

\begin{figure}[h]
  \centering
\begin{verbatim}
const int N=30;

// function that takes long time to analyze
int h(int y){
  int i,c=0;
  for (int j=0;j<N;j++){
    i = input();
    if (i == c)
      c++;
  }

  return c;
}

// some library call
void call(int i){
  if (i<0){
    printf("Goal unreach\n");
  } else {
    printf("Goal two\n");
  }
}

int main(void) {
  int i,j,r=0;
  i = 5;
  while(i>0){
    int tmp = h(i);
    if (tmp > r){
      r = tmp;
    } 
    i--;
  }
  call(r);  
}

\end{verbatim} 
  \caption{The ``unreach'' example.}
\label{fig:unreach}
\end{figure}

\end{document}